\documentclass[review,12pt]{elsarticle}
\usepackage{graphicx}
\usepackage{amssymb,amsmath,amsthm}
\usepackage{setspace}
\usepackage{color}
\usepackage{url}
\newtheorem{lem}{Lemma} % Assume a plain lemma style
\newtheorem{thm}{Theorem} % Assume a plain theorem style
\newtheorem{dfn}{Definition}
\journal{Discrete Applied Mathematics}
\begin{document}
\begin{frontmatter}
\title{Sensitivity of the Performance of a  Simple Exchange Model to its Topology\footnote{\color{red}This manuscript has been authored by Sandia Corporation under Contract No. DE-AC04-94AL85000 with the U.S. Department of Energy.  The United States Government retains and the publisher, by accepting the article for publication, acknowledges that the United States Government retains a non-exclusive, paid-up, irrevocable, world-wide license to publish or reproduce the published form of this manuscript, or allow others to do so, for United States Government purposes.}
}
\author[Sandia]{Vitus J. Leung}
\ead{vjleung@sandia.gov}
\author[Sandia]{Randall A. LaViolette}
\ead{ralavio@sandia.gov}
\address[Sandia]{Sandia National Laboratories, P.O. Box 5800, Albuquerque, NM, 87185 USA}

\begin{abstract}
We study a simple exchange model in which price is fixed and the amount of a good transferred between actors depends only on the actors' respective budgets and the existence of a link between transacting actors. The model induces a simply-connected but possibly multi-component bipartite graph. A trading session on a fixed graph consists of a sequence of exchanges between connected buyers and sellers until no more exchanges are possible. We deem a trading session ``feasible'' if all of the buyers satisfy their respective demands. If all trading sessions are feasible the graph is deemed ``successful'', otherwise the feasibility of a trading session depends on the order of the sequence of exchanges. We demonstrate that topology is important for the success of trading sessions on graphs. In particular, for the case that supply equals demand for each component of the graph, we prove that the graph is successful if and only if the graph consists of components each of which are complete bipartite. For the case that supply exceeds demand, we prove that the other topologies also can be made successful but with finite reserve (i.e., excess supply) requirements that may grow proportional to the number of buyers. Finally, with computations for a small instance of the model, we provide an example of the wide range of performance in which only the connectivity varies. These results taken together place limits on the improvements in performance that can be expected from proposals to increase the connectivity of sparse exchange networks.
\end{abstract}
\end{frontmatter}

\section{Introduction}
\label{sec:intro}
Networked infrastructures are designed to efficiently deliver goods between actors; furthermore, they are designed to continue functioning even if some components of the network fail. In most such infrastructures, suppliers maintain a reserve against a range of likely demand scenarios. One class of strategies to reduce such reserve requirements, and central to Smart Grid\cite{Kieseling,KieWil} proposals, for example, employ a dramatic increase in the connectivity of the network and the exchanges that take place on them\cite{MVM+08}. On the other hand, an increase in network connectivity may under some circumstances degrade rather than improve network performance, as illustrated by, e.g., the venerable Braess paradox and its variations\cite{NetworkStability,Hagstrom2001,Kameda2000,Korilis1995,Penchina2003,Roughgarden2002,Roughgarden2006}. Therefore we want to assess the impact of those proposed upgrades to networked infrastructures that increase its connectivity between actors. As a first step we study a simple model of exchanges between non-cooperative actors. In particular, we will show that it is easy to generate Braess-like paradoxes wherein the ability to meet all demands is degraded by nothing more than increasing the number of links. 

\section{Description of the Exchange Model}
\label{sec:model}

\subsection{Trading between a buyer and a seller}
\label{ssec:trading}
\begin{figure}[p]
\centering
\includegraphics[width=5cm]{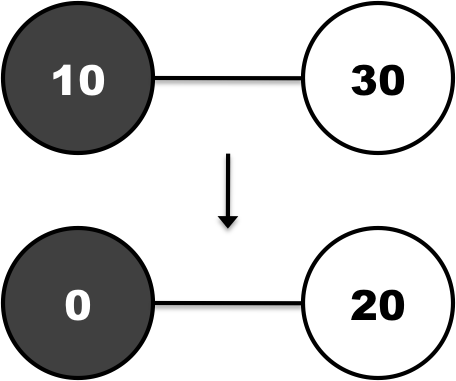}
\caption{An example of an exchange before (top) and after (bottom) the trade. Note that no further exchange is allowed on this link.}
\label{trade}
\end{figure}

First we consider an unsupervised bilateral trade between a buyer $b$ with demand $D_b$ and a seller $s$ with supply $S_s$; alternatively we may consider the exchange of two goods with ``supply'' as one good and ``demand'' as the other good. We assume that the trade exchanges supply for demand at a fixed unit exchange rate (price). The roles of seller and buyer are fixed, e.g., buyers do not become sellers. We signify the access of buyers to sellers with a link between the two actors. Trading is not optional; a trade between a buyer and seller must occur when (a) there exists a link between the buyer and the seller (b) the buyer has positive demand (c) the seller has positive supply. The amount traded is the maximum that can be traded given the available supply and demand (but see the end of \ref{ssec:statement} where we relax this requirement). Therefore at least one actor's supply or demand is always reduced to zero, i.e., if the buyer with demand $D_b = \delta$ trades with seller with supply $S_s = \sigma$, the result of the trade will be that the buyer is left with $\max (0,\delta - \sigma)$ and the seller is left with $\max (0,\sigma - \delta)$ (e.g., see Figure \ref{trade}). This simple budget-constrained exchange model\cite{exchange,LES+09} deviates from both other exchange models\cite{NIPS2005,EC07} and standard assumptions of economics\cite{MICRO:2001a} because we abandoned the classical concept of bi-modal traders by instead fixing an agent as either a buyer or a seller and we imposed a more restrictive specification of trading preferences; see \cite{LES+09} for a full discussion). 

\subsection{Trading between many buyers and sellers}
\label{ssec:many_trades}
The exchange model\cite{exchange} employed here then consists of a fixed set of $N_b$ buyers (with total demand ${\cal D}$) and $N_s$ sellers (with total supply ${\cal S}$). Throughout we will assume $\cal S = D$. The $L$ links between buyers and sellers induce a bipartite graph (see Figure \ref{graph}); e.g., the case in which all buyers are accessible to all sellers is the complete bipartite graph $K_{N_b,N_s}$. 

We define a trading session:
\begin{dfn}
A trading session on the graph consists of one of the (up to $L!$) possible sequences of all possible trades on the graph of $L$ links. 
\label{def:session}
\end{dfn}
Because each link trades at most once, it may happen that a trade never occurs over a link because a buyer or seller has been depleted by trades on other links. 
It may therefore happen that a demand remains unmet at the end of the trading session. Therefore we define ``feasible'' and ``successful'' as follows:
\begin{dfn}
A trading session is feasible if it reduces all demands to zero, otherwise it is infeasible.
\label{def:infeasible}
\end{dfn}
\begin{dfn}
A graph is successful if all possible trading sessions are feasible.
\label{def:successful}
\end{dfn}

\subsection{Statement of the problem}
\label{ssec:problem}
The complete bipartite graph $K_{N_b,N_s}$ is apparently successful according to Def. \ref{def:successful} if supply equals demand. On the other hand, inspection of Figure \ref{graph} shows that there is also a minimally connected multi-component bipartite graph that is successful. We ask, given a set of initial demands and initial supplies, which graphs are successful? We suspect that for some graphs there would be some orderings of the exchanges in which at least one demand would not be met at the end of the session: indeed we show in the next section that it is necessary and sufficient for each component (for which, within that component, supply equals demand) to be complete bipartite in order to be successful.

\section{Results}
\label{sec:results}
\subsection{Statement and Proof of the Theorem}
\label{ssec:statement}
Here we characterize the topology of successful (order-independent) graphs.
First we state and prove a useful lemma.
Then we state and prove the main theorem.
\begin{lem}
If the removal of the endpoints of a link in a bipartite component separates
the component into multiple components that are complete bipartite,
there exists a link in the component with endpoints that do not separate the
component into multiple components.
\label{lem1}
\end{lem}
\begin{proof}
Consider a link $l$ in a bipartite component with endpoints whose removal
separates the component into multiple components that are complete bipartite.
If any of the resulting components are single links, any one of them would be a
link with endpoints that do not separate the component into multiple
components.
If no components are single links, any link that does not contain the only
connections to $l$ for its component would be a link with endpoints that do
not separate the component into multiple components.
\end{proof}
\begin{thm}
Given that supply equals demand, the demands are reduced to zero at the end of every trading session iff each component (for which, within that component, supply equals demand) is complete bipartite.
\label{thm1}
\end{thm}
\begin{proof}[Sufficiency]
Assume for contradiction that a demand may not be reduced to zero at the end of a trading session when each component is complete bipartite and supply equals demand within each component.
We construct an example of this.
In our example, each component is complete bipartite, supply equals demand within each component, and a demand $D_b$ in component $C$ is not met at the end of a session.
However, this is impossible.
If some demand remains in $C$, some supply remains in $C$.
Since $C$ is complete bipartite, there exists a link between the buyer $b$ with unmet demand and a seller with remaining supply.
Therefore, a trade can occur.
This contradicts that the trading session has ended.
Therefore, when each component is complete bipartite and supply equals demand within each component, the demands are reduced to zero at the end of every trading session.
\end{proof}
\begin{proof}[Necessity]
Without loss of generality, we prove necessity for a component $C$ by strong induction.
Obviously, supply must equal demand within a component.
Let $C_{N_b,N_s}$ be a bipartite component with $N_b$ buyers and $N_s$ sellers.
Let $G_{N_b,N_s}$ be a bipartite graph with $N_b$ buyers and $N_s$ sellers.
\begin{description}
\item[Base case]
Consider a configuration with a single buyer $b$ and a single seller $s$ connected by a link $l$ and $S_s$ equals $D_b$.
If we remove $l$, the resulting configuration cannot be successful.
Therefore, $G_{1,1}$ must be $K_{1,1}$ for the demand of $b$ to be reduced to zero at the end of the trading session.
\item[Induction hypothesis]
Assume for all $N_b < m$ and $N_s \le n$ or $N_b \le m$ and $N_s < n$, that $C_{N_b,N_s}$ must be $K_{N_b,N_s}$ and supply must equal demand for $C_{N_b,N_s}$ to be successful.
\item[Inductive step]
Consider a component $C_{m,n}$ with $L$ links and at least one link between
some buyer $b$ and some seller $s$ such that the removal of the actor or actors
that can no longer participate after a first trade between $b$ and $s$ does not
result in multiple components.
The existance of such a link is guaranteed by Lemma \ref{lem1} and the
induction hypothesis if $C_{m,n}$ is successful.
Consider a first trade between $b$ and $s$.
After this first trade, either $D_b$ or $S_s$ has been reduced to zero.
In any case, we can remove the actor or actors that can no longer participate.
According to the induction hypothesis, the remaining component must be complete bipartite to be successful.
Now restore the actor or actors and undo the first trade.
There are three cases.
\begin{enumerate}
\item
Only $D_b$ was reduced to zero by the first trade.
Consider $b$'s links to sellers.
If $b$ does not have links to all sellers, sequencing trades so that the sellers to which $b$ is linked are reduced as much as possible by buyers other than $b$ would leave $b$'s demand unmet at the end
of the session.
This is possible because the component without $b$ is complete bipartite,
all supplies are greater than zero, and supply equals demand.
Therefore $b$ must have links to all sellers in order to have its demand reduced to zero at the end of every trading session.
\item
Only $S_s$ was reduced to zero by the first trade.
Consider $s$'s links to buyers.
If $s$ does not have links to all buyers, sequencing trades so that the buyers to which $s$ is linked are reduced as much as possible by sellers other than $s$ would leave at least some of $s$'s supply
isolated at the end of the session, and at least one buyer does not have its demand reduced to zero.
This is possible because the component without $s$ is complete bipartite, all demands are greater than zero, and supply equals demand.
Therefore $s$ must have links to all buyers in order to have all demand reduced to zero at the end of every trading session.
\item
$D_b=S_s$.
One of $b$ or $s$ must have degree at least two or they would be a separate component.
Consider the one with lower degree.
Pick randomly in case of a tie.
There are two cases.
\begin{description}
\item[$b$ is picked]
If $b$ does not have links to all sellers, sequencing trades so that the sellers to which $b$ is linked are reduced as much as possible by buyers other than $b$ would leave $b$'s demand unmet at the end
of the session.
This is possible because $D_b=S_s$, $s$ has degree at least two,
the component without $b$ and $s$ is complete bipartite,
all supplies are greater than zero, and supply equals demand.
Therefore $b$ must have links to all supplies in order to have its demand reduced to zero at the end of every trading session.
Now consider $s$'s links to buyers as in Case 2 above.
\item[$s$ is picked]
If $s$ does not have links to all buyers, sequencing trades so that the buyers to which $s$ is linked are reduced as much as possible by sellers other than $s$ would leave at least some of $s$'s supply
isolated at the end of the session, and at least one buyer does not have its demand reduced to zero.
This is possible because $D_b=S_s$, $b$ has degree at least two, the component without $b$ and $s$ is complete bipartite, all demands are greater than zero, and supply equals demand.
Therefore $s$ must have links to all buyers in order to have all demand reduced to zero at the end of every trading session.
Now consider $b$'s links to sellers as in Case 1 above.
\end{description}
Therefore, both $b$ and $s$ must have links to all sellers and buyers, respectively, in order to have all demand reduced to zero at the end of every trading session.
\end{enumerate}
Therefore $G_{m,n}$ must be $K_{m,n}$ in order to have all demand reduced to zero at the end of every trading session.
\end{description}
\end{proof}
If we relax the requirement that the amount traded is the maximum that can be
traded given the available supply and demand, the theorem would still hold
because a trading session could still use that maximum amount or combine
successive trades to give the same result.

\subsection{Enumeration}
\label{ssec:enumeration}
\begin{figure}[p]
\centering
\includegraphics[width=5cm]{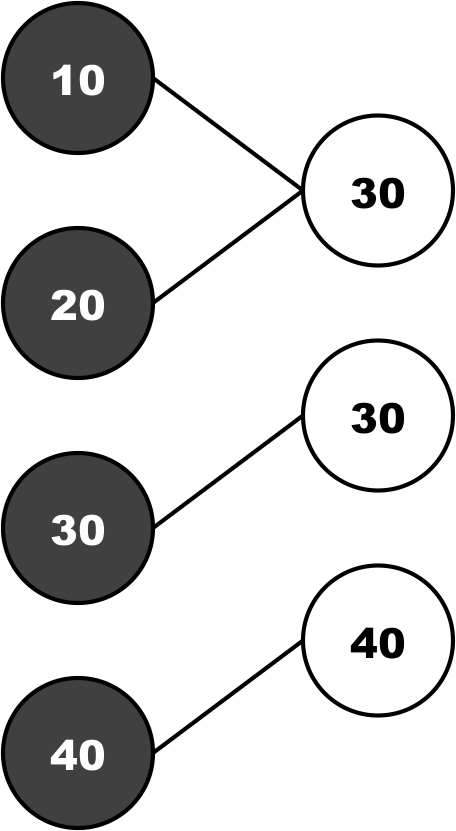}
\caption{An example of a exchange network at the beginning of a trading session. In this example, the demand will be met regardless of the order in which exchanges (specified by the links) occur.}
\label{graph}
\end{figure}

\begin{figure}[p]
\centering
\includegraphics[width=12cm]{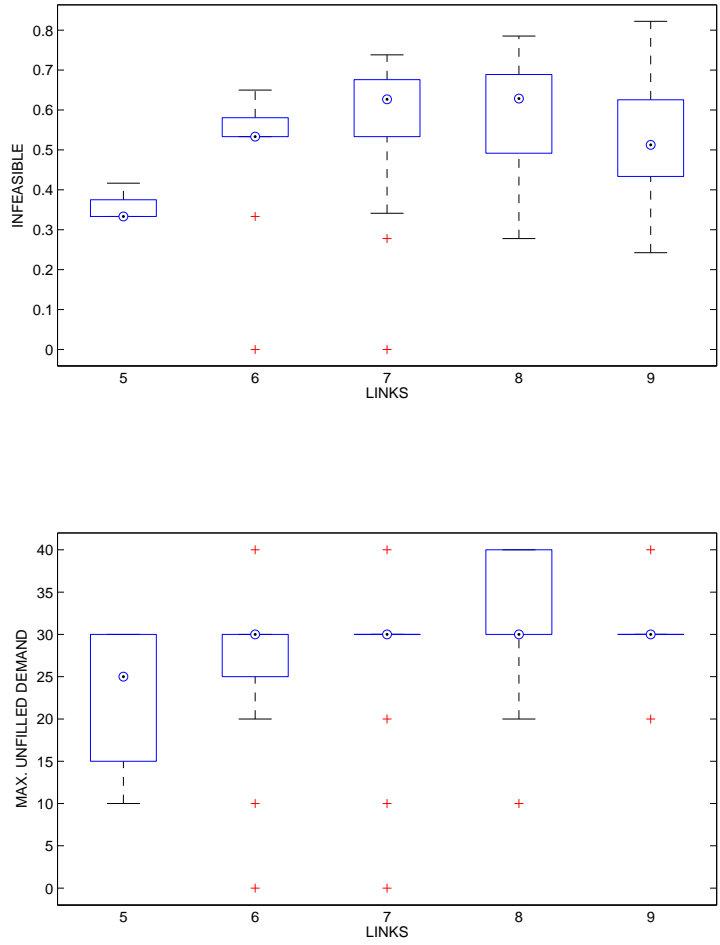}
\caption{Boxplots for results of the numerical evaluation of the fraction of infeasible orderings (top) and maximum unfilled demand (bottom) as a function of links, with initial conditions and the first four links fixed as in Figure \ref{graph}. The target is placed at the median, the top and bottom of the boxes correspond to the $75^{th}$ and $25^{th}$ percentiles, respectively. The crosses mark the outliers.}
\label{results}
\end{figure}

Here we study a special case (see Figure \ref{graph}) in order to directly enumerate the outcome of trading sessions. The graph in Figure \ref{graph} satisfies the conditions of Theorem \ref{thm1}. All subsequent graphs considered here have the backbone of the original four links shown in Figure \ref{graph}. We generate a new graph by adding $1 \le k \le 5$ links in every possible way to the original graph. For each $k$ there are $\binom{8}{k}$ graphs with $L=k+4$ links, each with $L!$ different trading sessions. We calculate the results of each trading session for each graph in order to calculate the fraction of infeasible trading sessions (Def. \ref{def:infeasible}) and the maximum demand left by infeasible sessions. The results are displayed in the top and bottom panels, respectively, of Figure \ref{results}. Of course we ensured in advance for this example that each component has total supply equal to total demand so that no configurations were trivially infeasible.
The addition of even one link to the graph of Figure \ref{graph} is enough to generate a substantial number of trading sessions that cannot reduce the demands to zero. Indeed none of the eight graphs with one extra link is successful. One of the 28 graphs with two links added (corresponding to $K_{2,1}\cup K_{2,2}$) is  successful but the median fraction of infeasible trading sessions on all 28 graphs is higher than the case of one link. The least successful graphs are those with four extra links. Subsequently, adding links improves the situation ($L=9$); finally, the case of $L=12$ (corresponding to $K_{4,3}$) is always successful.

The median maximum demand left over in unsuccessful graphs (with $\cal S = \cal D$)in the numerical experiments is $25-30\%$ of the initial demand.
The maximum
demand left over can actually be arbitrarily close to $100\%$ of the
initial demand as demonstrated by the lower bound in Section \ref{ssec:lb}.
This translates into large reserves that would be required by the sellers in order to meet demands on such graphs. 

\subsection{Statement and Proof of a Reserve Lower Bound}
\label{ssec:lb}
Here we characterize the reserves that would be required by sellers in order to
meet demands on unsuccessful graphs in the worst case.
\begin{thm}
The reserves that would be required by sellers in order to meet
demands on unsuccessful graphs can be proportional to $N_b$, the number of
buyers.
\label{thm2}
\end{thm}
\begin{proof}
Consider a successful graph consisting of $N_b$ buyers and $N_s=N_b$
sellers.  Each buyer has demand $D$ and each supplier has supply $S=D$.  Each
seller $s_i$ is connected to a buyer $b_i$.  Now create an unsuccessful graph
by adding links between seller $s_1$ and buyers $b_2, b_3, b_4, ..., b_{N_b}$.
The supply for $s_1$ would have to be increased to $N_b\cdot D$ in order for this
graph to be successful.
\end{proof}

\subsection{Computing the Maximum Reserve Requirement}
Here we give a nonlinear program to calculate the maximum demand left unmet in
an unsuccessful graph.
Let $l_{ij}$ equal one if the link between $b_i$ and $s_j$ exists and zero
otherwise.
Then let $t_{ij}\geq 0$ be the amount transacted over $l_{ij}$.
If $u_i\geq 0$ is the unmet demand at $b_i$,
the maximum unmet demand can be computed by the nonlinear program in Figure
\ref{nlp}.
Note that the addition of reserves can result in the need for still more
reserves as demonstrated in the lower bound in Section \ref{ssec:lb}.

By removing the objective and constraints that contain $u_i$ and changing the
inequality $$\sum_jt_{ij}\leq D_i,\forall i$$ to the equality
$$\sum_jt_{ij}=D_i,\forall i,$$ the nonlinear program in Figure \ref{nlp}
becomes the linear program in Figure \ref{lp} which can be used to determine
whether a graph is feasible.

\section{Conclusions}

For this simple exchange model on a bipartite graph representing transactions between $N_b$ buyers and $N_s$ sellers, we have shown that the sellers, with supply equal to demand, can satisfy demand for any sequence of transactions if and only if each component is complete bipartite. For a network of many small components, this requirement is not especially demanding. On the other hand, if the network is for some other reason required to consist of only one component, then there would exist some sequence of transactions for which sellers could not meet demand unless all $N_b \cdot N_s$ links were created or unless the initial supply had been supplemented by reserves.

We have a nonlinear program (Figure \ref{nlp}) to calculate the maximum demand left over in
an unsuccessful graph.
Nevertheless, the addition of reserves can result in the need for still
more reserves as demonstrated in the lower bound in Section \ref{ssec:lb}.

We have not been concerned here with the question of how links arise in the first place, i.e., what drives more {\em vs.} less links besides the minimal requirement to allow an exchange between a buyer and a seller. In the context of markets, the formation of the links in the first place would seem to require a model (e.g., \cite{NIPS2005}) in which prices are discovered, e.g., through auctions or brokers, which (as discussed in more detail in \cite{LES+09}) are left out of this model. In some contexts\cite{Gradwohl}, increasing the number of links might lower the price of exchanges for some actors, or, policies such as the SmartGrid\cite{Kieseling}, might for other reasons require many links. If the only penalty for adding links were the one-time cost of installing the links themselves, it would be unsurprising that the benefits of additional connectivity would, in time, outweigh its expense. Nevertheless our results provide another potential penalty for additional links between buyers and sellers that should be included in the cost-benefit calculation.

While we expect that most proposals to reduce reserve requirements would plan more than merely increased network connectivity, we have focused this study of the simple case in which only the connectivity has increased. Therefore we don't expect to apply these results directly to any particular proposal but instead expect these results to be only a part of a systematic evaluation of such proposals. Nevertheless, our results suggest that the mere increase in connectivity of a sparse exchange network will be problematic unless the details of the topology are considered explicitly. In particular, as both theory and computations suggest, there is a range of connectivity in these networks that impose reserve requirements that might cancel out the advantages that more links would otherwise provide.

\section*{Acknowledgements}
We thank our colleagues Bob Carr, Ben Cook, Jonathan Eckstein (RUTCOR), Hamilton Link, Verne Loose, Ojas Parekh, and Jason Stamp for helpful discussions and references. The work was supported in part by the Sandia National Laboratories Laboratory Directed Research and Development Program. Sandia National Laboratories is a multi-program laboratory operated by Sandia Corporation, a Lockheed Martin Company for the United States Department of Energy's National Nuclear Security Administration under contract DE-AC04-94AL85000.

\section*{Figures}

\begin{figure}[p]
$$\max\sum_iu_i,$$
subject to:
\begin{align*}
t_{ij}&\leq D_i\cdot l_{ij},\forall i,\forall j;\\
t_{ij}&\leq l_{ij}\cdot S_j,\forall i,\forall j;\\
\sum_jt_{ij}&\leq D_i,\forall i;\\
\sum_it_{ij}&\leq S_j,\forall j;\\
u_i&=D_i-\sum_jt_{ij},\forall i;\\
\left\{\sum_j\left[ l_{ij}\cdot \left( S_j-\sum_it_{ij}\right)\right]\right\}
 \cdot u_i&=0,\forall i.\\
\end{align*}
\caption{Nonlinear program to compute the maximum unmet demand for an
unsuccessful graph.}
\label{nlp}
\end{figure}

\begin{figure}[p]
\begin{align*}
t_{ij}&\leq D_i\cdot l_{ij},\forall i,\forall j;\\
t_{ij}&\leq l_{ij}\cdot S_j,\forall i,\forall j;\\
\sum_jt_{ij}&= D_i,\forall i;\\
\sum_it_{ij}&\leq S_j,\forall j;\\
\end{align*}
\caption{Linear program (constraint system) to determine the feasibility of a graph.}
\label{lp}
\end{figure}

\pagebreak

\bibliographystyle{elsarticle-num}
\bibliography{bib}

\begin{thebibliography}{10}
\expandafter\ifx\csname url\endcsname\relax
  \def\url#1{\texttt{#1}}\fi
\expandafter\ifx\csname urlprefix\endcsname\relax\def\urlprefix{URL }\fi
\expandafter\ifx\csname href\endcsname\relax
  \def\href#1#2{#2} \def\path#1{#1}\fi

\bibitem{Kieseling}
L.~Kiesling,
  \href{http://knowledgeproblem.com/2009/03/03/a-smart-grid-is-a-transactive-g%
rid-part-2-of-5/}{A smart grid is a transactive grid} (2009).
\newline\urlprefix\url{http://knowledgeproblem.com/2009/03/03/a-smart-grid-is-%
a-transactive-grid-part-2-of-5/}

\bibitem{KieWil}
L.~Kiesling, B.~J. Wilson, An experimental analysis of the effects of automated
  mitigation procedures on investment and prices in wholesale electricity
  markets, Journal of Regulatory Economics 31~(3) (2007) 313--334.

\bibitem{MVM+08}
M.~McGranaghan, D.~Von~Dollen, P.~Myrda, E.~Gunther, {Utility Experience with
  Developing a Smart Grid Roadmap}, in: {2008 IEEE POWER \& ENERGY SOCIETY
  GENERAL MEETING, VOLS 1-11}, {2008}, pp. {1193--1197}, {General Meeting of
  the IEEE-Power-and-Energy-Society, Pittsburgh, PA, JUL 20-24, 2008}.

\bibitem{NetworkStability}
E.~Anshelevich, A.~Dasgupta, J.~Kleinberg, E.~Tardos, T.~Wexler,
  T.~Roughgarden, {THE PRICE OF STABILITY FOR NETWORK DESIGN WITH FAIR COST
  ALLOCATION}, {SIAM JOURNAL ON COMPUTING} {38}~({4}) ({2008}) {1602--1623}.
\newblock \href {http://dx.doi.org/{10.1137/070680096}}
  {\path{doi:{10.1137/070680096}}}.

\bibitem{Hagstrom2001}
J.~N. Hagstrom, R.~A. Abrams, Characterizing braess's paradox for traffic
  networks, in: 2001 IEEE INTELLIGENT TRANSPORTATION SYSTEMS - PROCEEDINGS,
  {2001}, pp. {836--841}, {IEEE Intelligent Transportation Systems Conference
  (ITSC'01), Oakland, CA, AUG 25-29, 2001}.

\bibitem{Kameda2000}
H.~Kameda, E.~Altman, T.~Kozawa, Y.~Hosokawa, Braess-like paradoxes in
  distributed computer systems, IEEE Transactions on Automatic Control 45~(9)
  (2000) 1687--1691.

\bibitem{Korilis1995}
Y.~A. Korilis, A.~A. Lazar, A.~Orda, {Architecting Noncooperative Networks},
  IEEE Journal on Selected Areas in Communications 13~(7) (1995) 1241--1251.

\bibitem{Penchina2003}
C.~M. Penchina, L.~J. Penchina, {The Braess Paradox in mechanical, traffic, and
  other networks}, American Journal of Physics 71~(5) (2003) 479--482.

\bibitem{Roughgarden2002}
T.~Roughgarden, E.~Tardos, {How bad is selfish routing?}, JOURNAL OF THE ACM
  {49}~({2}) ({2002}) {236--259}.

\bibitem{Roughgarden2006}
T.~Roughgarden, {On the severity of Braess's Paradox: Designing networks for
  selfish users is hard}, {JOURNAL OF COMPUTER AND SYSTEM SCIENCES} {72}~({5})
  ({2006}) {922--953}, {42nd Annual IEEE Symposium on Foundations of Computer
  Science, Las Vegas, NV, OCT 14-17, 2001}.
\newblock \href {http://dx.doi.org/{10.1016/j.jcss.2005.05.009}}
  {\path{doi:{10.1016/j.jcss.2005.05.009}}}.

\bibitem{exchange}
R.~A. LaViolette, L.~A. Ellebracht, C.~J. Gieseler, Limits on relief through
  constrained exchange on random graphs, Physica A 383 (2007) 671--676.

\bibitem{LES+09}
R.~A. LaViolette, L.~A. Ellebracht, K.~L. Stamber, C.~J. Gieseler, B.~K. Cook,
  \href{http://arxiv.org/abs/0905.2366}{Emergence of price divergence in a
  model short-term electric power market}, preprint (2009).
\newline\urlprefix\url{http://arxiv.org/abs/0905.2366}

\bibitem{NIPS2005}
S.~M. {Kakade}, M.~{Kearns}, L.~E. {Ortiz}, R.~{Pemantle}, S.~{Suri}, Economic
  properties of social networks, in: L.~K. Saul, Y.~Weiss, L.~Bottou (Eds.),
  Advances in Neural Information Processing Systems 17, MIT Press, Cambridge,
  MA, 2005, pp. 633--640.

\bibitem{EC07}
L.~Blume, D.~Easley, J.~Kleinberg, E.~Tardos, Trading networks with
  price-setting agents, in: Eighth ACM Conference on Electronic Commerce
  (EC07), ACM, 2007.

\bibitem{MICRO:2001a}
R.~S. Pindyck, D.~L. Rubinfeld, Microeconomics, 5th Edition, Prentice-Hall,
  Upper Saddle River, NJ, 2001.

\bibitem{Gradwohl}
R.~Gradwohl, {Price variation in a bipartite exchange network}, in: {Monien, B
  and Schroeder, UP} (Ed.), {ALGORITHMIC GAME THEORY, PROCEEDINGS}, Vol. {4997}
  of {LECTURE NOTES IN COMPUTER SCIENCE}, {SPRINGER-VERLAG BERLIN},
  {HEIDELBERGER PLATZ 3, D-14197 BERLIN, GERMANY}, {2008}, pp. {109--120}, {1st
  International Symposium on Algorithmic Game Theory, Paderborn, GERMANY, APR
  30-MAY 02, 2008}.

\end{thebibliography}

\end{document}